# Unabridged phase diagram for single-phased FeSe$_x$Te$_{1-x}$ thin films

Jincheng Zhuang[1,2], Wai Kong Yeoh[2,3,4,*], Xiangyuan Cui[3,4], Xun Xu[2], Yi Du[2], Zhixiang Shi[1,*], Simon P. Ringer[3,4], Xiaolin Wang[2] & Shi Xue Dou[2]

[1]Department of Physics and Key Laboratory of MEMS of the Ministry of Education, Southeast University, Nanjing 211189, People's Republic of China, [2]Institute for Superconducting and Electronic Materials, University of Wollongong, North Wollongong, New South Wales 2500, Australia, [3]Australian Centre for Microscopy and Microanalysis, University of Sydney, Sydney, New South Wales 2006, Australia, [4]School of Aerospace, Mechanical and Mechatronic Engineering, University of Sydney, New South Wales 2006, Australia.

*Correspondence and requests for materials should be addressed to W. K. Y. (waikong.yeoh@sydney.edu.au) or Z. X. S. (zxshi@seu.edu.cn)

**A complete phase diagram and its corresponding physical properties are essential prerequisites to understand the underlying mechanism of iron-based superconductivity. For the structurally simplest 11 (FeSeTe) system, earlier attempts using bulk samples have not been able to do so due to the fabrication difficulties. Here, thin FeSe$_x$Te$_{1-x}$ films with the Se content covering the full range ($0 \leq x \leq 1$) were fabricated by using pulsed laser deposition method. Crystal structure analysis shows that all films retain the tetragonal structure in room temperature. Significantly, the highest superconducting transition temperature ($T_C$ = 20 K) occurs in the newly discovered domain, *i.e.*, $0.6 \leq x \leq 0.8$. The single-phased superconducting dome for the full Se doping range is the first of its kind in iron chalcogenide superconductors. Our results present a new avenue to explore novel physics as well as to optimize superconductors.**

Shortly after the discovery of iron-based superconductors, in the form of LaFeAsO$_{1-x}$F$_x$, with a critical transition temperature ($T_C$) of 26 K[1], superconductivity was observed in PbO type iron chalcogenides (11 system)[2]. Due to its simple crystal structure, composed of a stack of superconducting Fe$_2$Ch$_2$ (Ch = Se, Te) layers along the c-axis, as well as its rich phase diagram[3], the 11 system has attracted tremendous interest in exploring the mechanism of high temperature superconductivity. Although the $T_C$ of FeSe is as low as 8 K, it can be substantially enhanced up to 15 K with partial substitution of Te for Se at $x = 0.5$ for both polycrystalline[3] and single crystal samples[4], or to 37 K under high pressure[5,6]. Moreover, under the influence of strain effects induced by the lattice mismatch between the film and substrate[7], or by Fe vacancies[8], a maximum $T_C$ of 21 K can be obtained in FeSe$_{0.5}$Te$_{0.5}$ films. The recent discovery of a record high $T_C$ of 65 K in monolayer FeSe films [9,10] led to the exciting perspective that its $T_C$ can be as high as the liquid nitrogen boiling temperature. Nevertheless, a detailed phase diagram for the 11 system is yet to be established[3,4]; in particular, the physical properties of samples with high Se content, $0.6 \leq x \leq 0.8$, are still unclear. One of the major reasons for such uncertainty is the lack of samples with the right stoichiometric ratios. Polycrystalline samples with Se concentrations in the range of $0.6 \leq x \leq 0.8$ are found to have multiple-phases[4], while pure single crystal samples have never been fabricated. As a result, the FeSe$_{0.5}$Te$_{0.5}$ ($x = 0.5$) sample is regarded to have the maximum $T_C$ so far, based on the incomplete phase diagram, which might be misleading. Here, we have overcome the phase separation issues and successfully fabricated single-phased 11 films for $0.6 \leq x \leq 0.8$ via the pulsed laser deposition (PLD) method. By characterizing the structural and superconducting properties for the full range of FeSe$_x$Te$_{1-x}$ thin films, for the first time, we were able to construct the complete phase diagram. In particular, we uncovered a higher $T_C$ phase can be obtained in $x = 0.6$-$0.8$ samples, in contrast to the previous assumption of $x = 0.5$. Also interestingly, the $x = 0.6$ sample exhibits the maximum upper critical magnetic field

($H_{c2}$) compared to the other samples. This new phase diagram would now be strikingly critical for potential superconductor application and a better understanding of the superconducting mechanism.

**Results**

As reference systems, we first start with bulk samples. Figure 1(a) shows the XRD patterns for the targets (bulk) with different Se concentrations, where all the XRD peaks were normalized by the values of the intensity of the respective (101) peaks for direct comparison. From the XRD analysis, all samples seem to be well-developed single phase and can be indexed as PbO tetragonal structure except in the $0.6 \leq x \leq 0.8$ region. In that particular region, all the peaks split into two groups, indicating the coexistence of two structural phases, denoted as phase A and phase B, respectively, as shown in Figure 1(a). Phase A exists for $0 \leq x \leq 0.8$, while phase B emerges in the region $0.6 \leq x \leq 1$. Figure 1(b) displays the calculated values of the lattice parameters as a function of Se content in the bulk samples. For phase A, both the lattice parameters $a$ and $c$ decrease remarkably with increasing Se concentration, while for phase B, the lattice parameters change only slightly with variation of $x$. These results are consistent with a previous report[4], indicating that the structure of FeSe is different from the structure of FeTe, even though both of them have a similar tetragonal structure. In the dual phase region of $0.6 \leq x \leq 0.8$, both $a$ and $c$ in the A and B phases show almost no dependence on $x$, indicating that either A or B possesses the same stoichiometric composition among the three samples. The only variation among these three samples ($x = 0.6, 0.7,$ and $0.8$) is the component ratio between phase A and B, which can be identified by the changes in the diffraction peak intensity in Figure 1(a). Quantitative results on the peak intensity, as shown in Supplementary Fig. S1, demonstrate the linear variation of the A or B component as a function of Se content. The variations in the lattice parameters and the linear variation of the component ratio with changing $x$ imply that there is no single phase in the bulk samples for the $0.6 \leq x \leq 0.8$ region.

In sharp comparison, there is no phase separation for the peak reflection of all thin films samples. Figure 1(c) displays the XRD patterns for thin films normalized by the intensity of their respective (001) peaks, where all films are denoted by the nominal composition of their targets. Only the (00$l$) reflections of thin films and of substrates can be detected, implying the out-of-plane orientation of these films. Interestingly, all the peaks show a systematic shift with changing $x$. Figure 1(d) displays a linear relationship between lattice parameter $c$ and the Se content $x$, indicating that the Se contents in the thin films are close to the nominal values and that single phase for $0.6 \leq x \leq 0.8$ is actually formed in the thin film state. This is crucial for material characterization as well as for phase diagram construction.

Figure 2(a) shows the electronic resistivity versus temperature ($R$-$T$) curves for all film samples normalized by their respective resistance at 300 K. For the un-doped FeTe film, a kink around 70 K is observed, corresponding to the antiferromagnetic (AFM) transition accompanied by a structural transition[3,4,11]. The kink disappears when the Se content increases up to $x = 0.2$, which is similar to that found in polycrystalline samples[4] and single crystal[3]. With further Se doping, the $R$-$T$ curve in the normal state becomes more metallic, and $T_C$ increases from 10 K at $x = 0.2$ up to around 20 K for $x$ in the range of 0.6 to 0.8, as shown in Figure 2(b). It should be noted that $T_C$ of the FeTe$_{0.5}$Se$_{0.5}$ film is around 16 K, consistent with the results of bulk[4] and single crystal samples[3]. There are a lot of works[7,8,12-15] indicating that strain, caused by the mismatch between thin films and substrates, plays a critical role in the emergency of higher $T_C$ (~19 K) compared with the bulk value. Nevertheless, with all the films being prepared on same substrate (CaF$_2$) in our study and a linear relationship between lattice parameter and Se concentration, the dominant factor for the $T_C$ variation is attributed to the Se doping rather than lattice mismatch.

Measurements of electronic resistivity as a function of temperature for all films, under different magnetic fields ranging from 0 T to 13 T parallel to the $c$ direction, were also performed. Figure 3(a) displays $R$-$T$ curves for FeSe$_{0.6}$Te$_{0.4}$ film samples at temperatures from 12 K to 22 K. All the other curves are displayed in Supplementary Fig. S2. Broadening of the resistivity transition in magnetic field is direct evidence of thermal fluctuation in a vortex system[16]. The width of the superconducting transition is calculated from the formula $\Delta T_C = T_C^{onset} - T_C^{zero}$, where $T_C^{onset}$ and $T_C^{zero}$ are determined by 90% and 10% of the normal state resistivity for all the films, respectively. The broadening of the transition width can be characterized by ($\Delta T_C(B) - \Delta T_C(0)$), where $\Delta T_C(B)$ is the transition width under the magnetic field. Figure 3(b) shows the broadening of the superconducting transition width as a function of magnetic field for all the superconducting films. It can be seen that samples of $0.6 \leq x \leq 0.8$ show higher slope gradient in comparison to other samples. In general, such a difference may reflect different vortex mechanism or superconducting order parameters between $T_C$ samples and other samples. The upper critical field ($H_{c2}$) is determined by 90% of normal state resistivity for all films. Figure 3(c) shows the dependence of the upper critical field for all superconducting films on the normalized temperature (by the respective $T_C$ values). It can be seen clearly that the highest $H_{c2}$ exists in the sample with $x = 0.6$. The values of $H_{c2}$ were obtained by using the Werthamer-Helfand-Hohenberg model[17], $-H_{c2}(0) = 0.67 T_C dH_{c2}/dT|T_C$. The estimated value of $H_{c2}(0)$ exceeds 100 T, which is expected to be useful for high-field applications.

Charge carrier population is one of the major parameters that are critical to high $T_C$ in Fe-based superconductors[18]. Therefore, we performed Hall measurements on three representative film samples, namely an under-doped sample (FeSe$_{0.3}$Te$_{0.7}$), an optimally-doped sample (FeSe$_{0.6}$Te$_{0.4}$), and an over-doped sample (FeSe$_{0.9}$Te$_{0.1}$). Figure 4(a-c) shows the transverse resistivity, $\rho_{xy}$, at different temperatures for $x = 0.3$, 0.6, and 0.9, respectively. All curves show linear-like behavior in the field measurement range. The Hall coefficient, $R_H = \rho_{xy}/B$, where $B$ is the magnetic flux density, is determined by the linear fitting of $\rho_{xy}$ curves. Figure 4(d) demonstrates that the $R_H$ of all three films is almost temperature independent from 80 K to room temperature. This phenomenon is similar to that in both films[8,19] and single crystal[20]. In the low temperature regime below 80 K, $R_H$ for all three films shows an obvious divergence with temperature. $R_H$ values for $x = 0.3$ and 0.9 show a noticeable upturn at low temperature, while $R_H$ for $x = 0.6$ starts to decrease at 60 K and finally changes sign from positive to negative before the superconducting transition. To determine the nature of charge carrier, the classical formula for the Hall coefficient of semiconductors in the presence of both electron- and hole-type carriers is used:[21]

$$R_H = \frac{1}{e} \frac{(\mu_h^2 n_h - \mu_e^2 n_e) + (\mu_h \mu_e)^2 B^2 (n_h - n_e)}{(\mu_e n_h + \mu_h n_e)^2 + (\mu_h \mu_e)^2 B^2 (n_h - n_e)^2} \qquad (1)$$

where $\mu_h$, $\mu_e$, $n_h$, and $n_e$ are hole mobility, electron mobility, hole density, and electron density, respectively. Eq. (1) predicts that $R_H = e^{-1}(\mu_h^2 n_h - \mu_e^2 n_e)/(\mu_e n_h + \mu_h n_e)^2$ when $B \to 0$, and $R_H = e^{-1} 1/(n_h - n_e)$ in the limit of $B = \infty$. Thus, the sign reversal has been claimed as the evidence for the multiband nature in the 11 system[22]. The positive $R_H$ values indicate that hole-type carriers are dominant for $x = 0.3$ and 0.9. On the other hand, the sign reversal of $R_H$ indicates that the population of electron carriers increase and become dominant for the $x = 0.6$ film. We believe that such the increase in electron carriers might be one of the primary origins for the higher $T_C$ in the $x = 0.6$ film. Our results clearly show the discrete variation of $n$ is not synchronous with continuously increasing Se concentration, indicating that the unusual charge carrier population is developed in the high $T_C$ region ($0.6 \leq x \leq 0.8$).

**Discussion**

According to conventional valency assumption, Se doping can be considered as isovalent element doping, and no charge carriers should be induced in this case. As a result, we think that local structural variation plays a critical role in driving the emergence of the unique charge carrier population, which has stimulated the highest $T_C$ in the region of $0.6 \leq x \leq 0.8$. The substitution of Se for Te is expected to induce a significant reduction in anion height[23,24] and to alter the magnetic orders in the FeSe$_x$Te$_{1-x}$ system[25]. When the anion height is reduced below a critical value, the magnetic ground state will switch from a bicollinear ($\pi$, 0) to collinear ($\pi$, $\pi$) pattern, which gives rise to superconducting pairing[3]. Although the long-range AFM order in FeTe is suppressed at $x = 0.3$, the short-range magnetic correlation is still present, which localizes the charge carriers and thus enhances $|R_H|$ at low temperature. This localization has been observed in 11 single crystals[26,27]. Furthermore, it has been reported that the dimensionality of the Fermi surface is quite sensitive to local structural properties, or the anion height[10,23], in iron-based superconductors. The reduction in anion height could, in turn, affect the density of states (DOS) at the Fermi energy ($E_F$) and the carrier density[28], by tuning the overlap between the Te 5$p$ (Se 4$p$) and Fe 3$d$ orbits. It should be noted that the sign reversal from positive to negative in $R_H$ ($x = 0.6$) should be considered as an intrinsic property for high $T_C$ samples in the 11 system, which has been found in high-quality thin film[19] and single crystal samples[29]. Our results thus highlight the role of structural parameters in the determination of charge carrier behaviour because both the AFM order and the modification of the Fermi surface show a strong relationship with anion height[23,24].

Collectively, our results enable us to construct the complete electronic phase diagram for the 11 film system, as shown in Figure 5. The AFM transition temperature, $T_N$, is determined by the position of the kink mentioned above, and $T_C$ is determined as the intersection point by extrapolation of the normal state resistivity and the superconducting transition, as shown in Figure 2. The previously reported $T_C$ values of polycrystalline[4,11] and single crystal[3] samples are also included for a comparative study. The AFM state FeTe is considered as the parent compound for 11 system. With the long range AFM being suppressed, superconductivity emerges through the Se doping with $T_C$ increasing gradually up to ~ 16 K for $x = 0.5$. The phase diagram displays a similar behavior as previous reports except for $x = 0.6$-$0.8$, where a higher $T_C$ was observed in the thin films compared to bulk samples. This results in an extended domain compared with bulk samples in the phase diagram, which proclaims the intrinsic properties of pure phase samples instead of the miscible phases. Together, these results highlight the intimate interplay between structural variation, charge carrier and magnetic interaction in Fe-based superconductors.

In summary, we have investigated the structural and transport properties of FeSe$_x$Te$_{1-x}$ film samples with $x$ ranging from 0 to 1. Single-phased film samples with high Se concentration ($0.6 \leq x \leq 0.8$), were successfully fabricated for the first time and possess the maximum $T_C$. Distinctive behavior of the charge carrier density could be evoked by the variation of structural parameters, which have a strong correlation with $T_C$. The electronic phase diagram containing the entire range of Se doping levels has been plotted, which revises the incomplete old phase diagram of 11 system.

**Methods**

**Sample preparation.** Polycrystalline pellets of FeTe$_{1-x}$Se$_x$ with nominal composition $x$ = 0, 0.1, 0.2, 0.3, 0.4, 0.5, 0.6, 0.7, 0.8, and 0.9 were fabricated as targets. Powders of Fe, Se, and Te, were mixed together in stoichiometric ratios and heated in an evacuated quartz tube at 850 ºC for 12 h. After the sintering, the mixture was reground, pelletized, and sintered in the evacuated quartz tube at 400 ºC for 6 h to make the target dense. The films were grown under vacuum conditions (~ 4 × 10$^{-4}$ Pa) by PLD using a Nd: YAG laser, as reported alsewhere[8]. Single crystal CaF$_2$ (100) with lattice parameter $a_0$ = 5.463 Å was selected as the substrate due to its non-oxide nature and the low mismatch between its lattice parameter ($a_0/\sqrt{2}$ = 3.863 Å) and the *a*-axis parameter of the film (around 3.8 Å). The deposition temperature was set at 450 ºC, and the laser energy was 200 mJ/pulse. The substrate-target distance was maintained at 4 cm, and the deposition time was the same for all samples.

**Measurements.** The thickness of the films, measured by scanning electron microscopy (SEM), was around 50 nm. The crystal structure and orientation of the films were characterized by X-ray diffraction (XRD) at room temperature. Electrical resistivity ($\rho$) and Hall measurements under different magnetic fields were carried out on a 14 T physical properties measurement system (PPMS).

1. Kamihara, Y., Watanabe, T., Hirano, M. & Hosono, H. Iron-based layered superconductor LaO$_{1-x}$F$_x$FeAs ($x$ = 0.05-0.12) with $T_C$ = 26 K. *J. Am. Chem. Soc.* **130**, 3296-3297 (2008).

2. Hsu, F. C. *et al.* Superconductivity in the PbO-type structure alpha-FeSe. *Proc. Natl. Acad. Sci. U.S.A.* **105**, 14262-14264 (2008).

3. Liu, T. J. *et al.* From (π, 0) magnetic order to superconductivity with (π, π) magnetic resonance in Fe$_{1.02}$Te$_{1-x}$Se$_x$. *Nat. Mater.* **9**, 718-720 (2010).

4. Fang, M. H. *et al.* Superconductivity close to magnetic instability in Fe(Se$_{1-x}$Te$_x$)$_{0.82}$. *Phys. Rev. B* **78**, 224503 (2008).

5. Medvedev, S. *et al.* Electronic and magnetic phase diagram of $\beta$-Fe$_{1.01}$Se with superconductivity at 36.7 K under pressure. *Nat. Mater.* **8**, 630 (2009).

6. Margadonna, S. *et al.* Pressure evolution of low-temperature crystal structure and bonding of the superconductor FeSe ($T_C$ = 37 K). *Phys. Rev. B* **80**, 064506 (2009).

7. Bellingeri, E. *et al.* $T_C$ = 21 K in epitaxial FeSe$_{0.5}$Te$_{0.5}$ thin film with biaxial compressive strain. *Appl. Phys. Lett.* **96**, 102512 (2010).

8. Zhuang, J. C. *et al.* Enhancement of transition temperature in Fe$_x$Se$_{0.5}$Te$_{0.5}$ film via iron vacancies. *Appl. Phys. Lett.* **104**, 262601 (2014).

9. He, S. L. *et al.* Phase diagram and electronic indication of high-temperature superconductivity at 65 K in single-layer FeSe films. *Nat. Mater.* **12**, 605-610 (2013).

10. Tan, S. Y. *et al.* Interface-induced superconductivity and strain-dependent spin density waves in FeSe/SrTiO$_3$ thin films. *Nat. Mater.* **12**, 634-640 (2013).

11. Yeh, K. W. *et al.* Tellurium substitution effect on superconductivity of the α-phase iron selenide. *EPL* **84**, 37002 (2008).

12. Si, W. D. *et al.* Enhanced superconducting transition temperature in FeSe$_{0.5}$Te$_{0.5}$ thin films. *Appl. Phys. Lett.* **95**, 052504 (2009).

13. Si, W. D. *et al.* Iron-chalcogenide FeSe$_{0.5}$Te$_{0.5}$ coated superconducting tapes for high field applications. *Appl. Phys. Lett.* **98**, 262509 (2011).

14. Si, W. D. *et al.* High current superconductivity in FeSe$_{0.5}$Te$_{0.5}$-coated conductors at 30 tesla. *Nat. Commun.* **4**, 1347 (2013).

15. Bellingeri, E. *et al.* Tunning of the superconducting properties of FeSe$_{0.5}$Te$_{0.5}$ thin films through the substrate effect. *Supercond. Sci. Technol.* **25**, 084022 (2012).

16. Maiorov, B. *et al.* Liquid vortex phase and strong *c*-axis pinning in low anisotropy BaCo$_x$Fe$_{2-x}$As$_2$ pnictide films. *Supercond. Sci. Technol.* **24**, 055007 (2011).

17. Werthamer, N. R., Helfand, E. & Hohenberg, P. C. Temperature and purity dependence of the superconducting critical field, $H_{c2}$. III. Electron Spin and Spin-Orbit Effects. *Phys. Rev.* **147**, 295 (1966).

18. Borisenko, S. Fewer atoms, more information. *Nat. Mater.* **12**, 600-601 (2013).

19. Tsukada, I. *et al.* Hall effect in superconducting Fe(Se$_{0.5}$Te$_{0.5}$) thin films. *Phys. Rev. B* **81**, 054515 (2010).

20. Sun, Y. et al. Multiband effects and possible Dirac fermions in Fe$_{1+y}$Te$_{0.6}$Se$_{0.4}$. *Phys. Rev. B* **89**, 144512 (2014).

21. Smith, R. A. *Semiconductors* (Cambridge University Press, Cambridge, UK, 1978).

22. Tsukada, I. *et al.* Hall effect of FeTe and Fe(Se$_{1-x}$Te$_x$) thin films. *Physica C* **471**, 625-629 (2011).

23. Subedi, A., Zhang, L. J., Singh, D. J. & Du, M. H. Density functional study of FeS, FeSe, and FeTe: Electronic structure, magnetism, phonons, and superconductivity. *Phys. Rev. B* **78**, 134514 (2008).

24. Mizuguchi, Y. *et al.* Anion height dependence of $T_C$ for the Fe-based superconductor. *Supercond. Sci. Technol.* **23**, 054013 (2010).

25. Moon, C. -Y. & Choi, H. J. Chalcogen-height dependent magnetic interactions and magnetic order switching in FeSe$_x$Te$_{1-x}$. *Phys. Rev. Lett.* **104**, 057003 (2010).

26. Sun, Y. *et al.* Dynamics and mechanism of oxygen annealing in Fe$_{1+y}$Te$_{0.6}$Se$_{0.4}$ single crystal. *Sci. Rep.* **4**, 4585 (2014).

27. Liu, T. J. *et al.* Charge-carrier localization induced by excess Fe in the superconductor Fe$_{1+y}$Te$_{1-x}$Se$_x$. *Phys. Rev. B* **80**, 174509 (2009).

28. Huang, S. X., Chien, C. L., Thampy, V. & Broholm, C. Control of tetrahedral coordination and superconductivity in FeSe$_{0.5}$Te$_{0.5}$ thin films. *Phys. Rev. Lett.* **104**, 217002 (2010).

29. Tropeano, M. *et al.* Transport and superconducting properties of Fe-based superconductors: a comparison between SmFeAsO$_{1-x}$F$_x$ and Fe$_{1+y}$Te$_{1-x}$Se$_x$. *Supercond. Sci. Technol.* **23**, 054001 (2010).

## Acknowledgements

This work was supported by the National Basic Research Program of China (973 Program, Grant No.2011CBA00105), National Science Foundation of China (Grant No. NSFC-U1432135), Scientific Research Foundation of Graduate School (Grant No. YBJJ1314) of Southeast University, and the Australian Research Council through Discovery Project (DP 120100095 and DP 140102581).

## Author contributions

W.K.Y, Z.S, S.P.R and S.X.D conceived and directed the research. Samples were prepared by J.Z and X.C. XRD data was collected and analysed by X.X and X.W. Y.D and J.Z contributed to superconducting and transport properties measurement. All authors discussed the results; J.Z and W.K.Y wrote the manuscript, with discussions mainly with X.C.

## Additional information

Competing financial interests: The authors declare no competing financial interests.

# Figure Captions

Figure 1 | (a) XRD patterns of FeSe$_x$Te$_{1-x}$ targets. The marks A and B stand for two phases coexisting in the targets for $0.6 \leq x \leq 0.8$. (c) XRD reflections of FeSe$_x$Te$_{1-x}$ thin films deposited on CaF$_2$ (100) substrate. "♦" marks the reflections of impurity phases in the substrate, identified by ICDD card. The corresponding calculated lattice parameters of the targets and thin films are shown in (b) and (d), respectively.

Figure 2 | (a) Electronic resistivity versus temperature (*R-T*) curves from 300 K to 2 K for all the films. (b) Enlarged view of *R-T* curves from 5 K to 25 K.

Figure 3 | (a) *R-T* curves from 22 K to 12 K under different magnetic fields up to 13 T parallel to the *c*-axis for FeSe$_{0.6}$Te$_{0.4}$ film. (b) Broadening of the width of the superconducting transition as a function of magnetic field for all superconducting films. The straight line is plotted only for guidance to the eye. (c) The upper critical field for all superconducting films depends on the temperature normalized by the respective $T_C$.

Figure 4 | Linear relationship between transverse resistivity and *B* at different temperatures for (a) *x* = 0.3, (b) *x* = 0.6, (c) *x* = 0.9. (d) Temperature dependence of Hall coefficient for the three films.

Figure 5 | The electronic phase diagram for FeSe$_x$Te$_{1-x}$ films ($0 \leq x \leq 1$) as a function of Se concentration. The $T_C$ of bulk and single crystal samples are also included for reference. The dashed line is used for the guidance to indicate the superconducting dome of bulk samples.

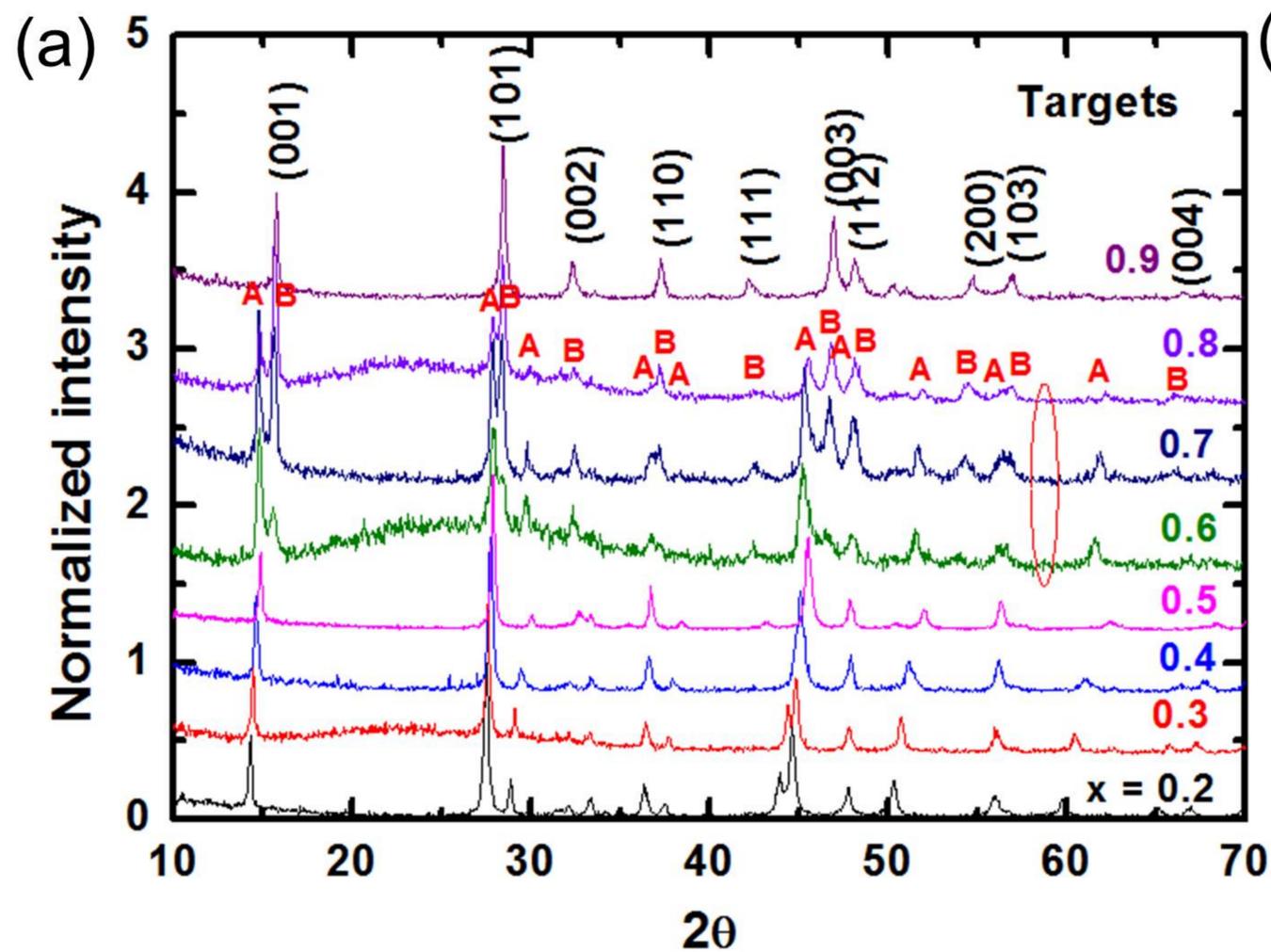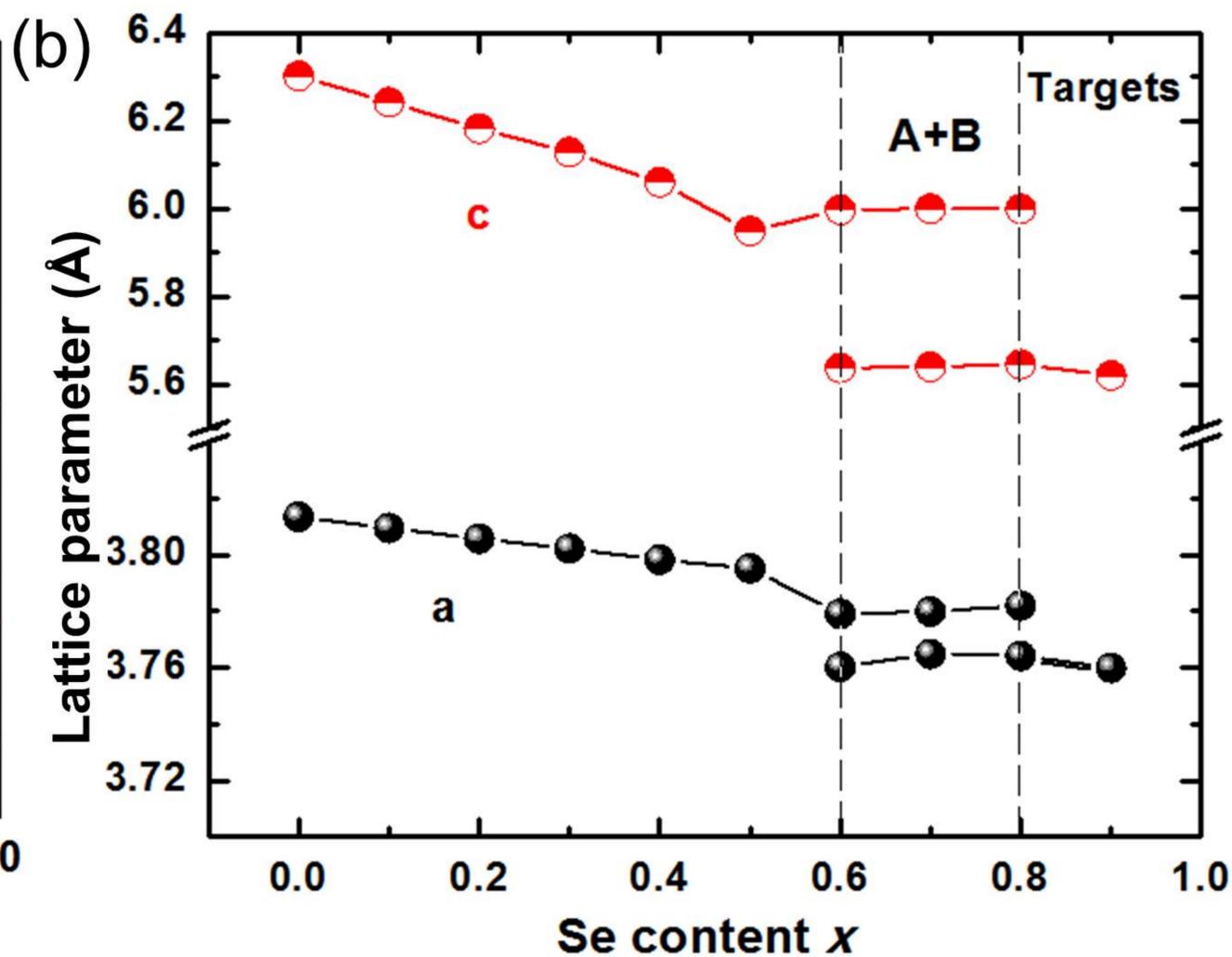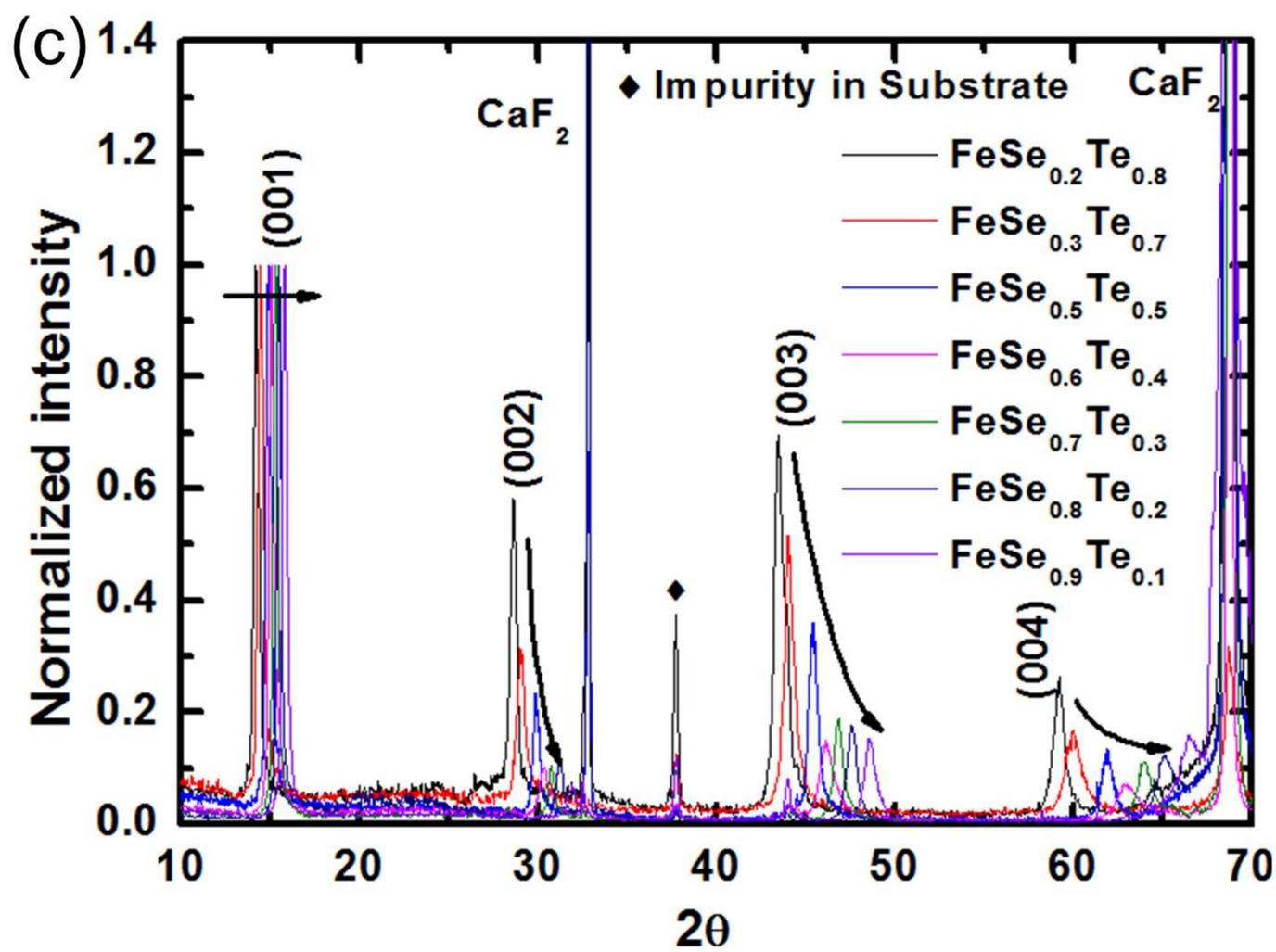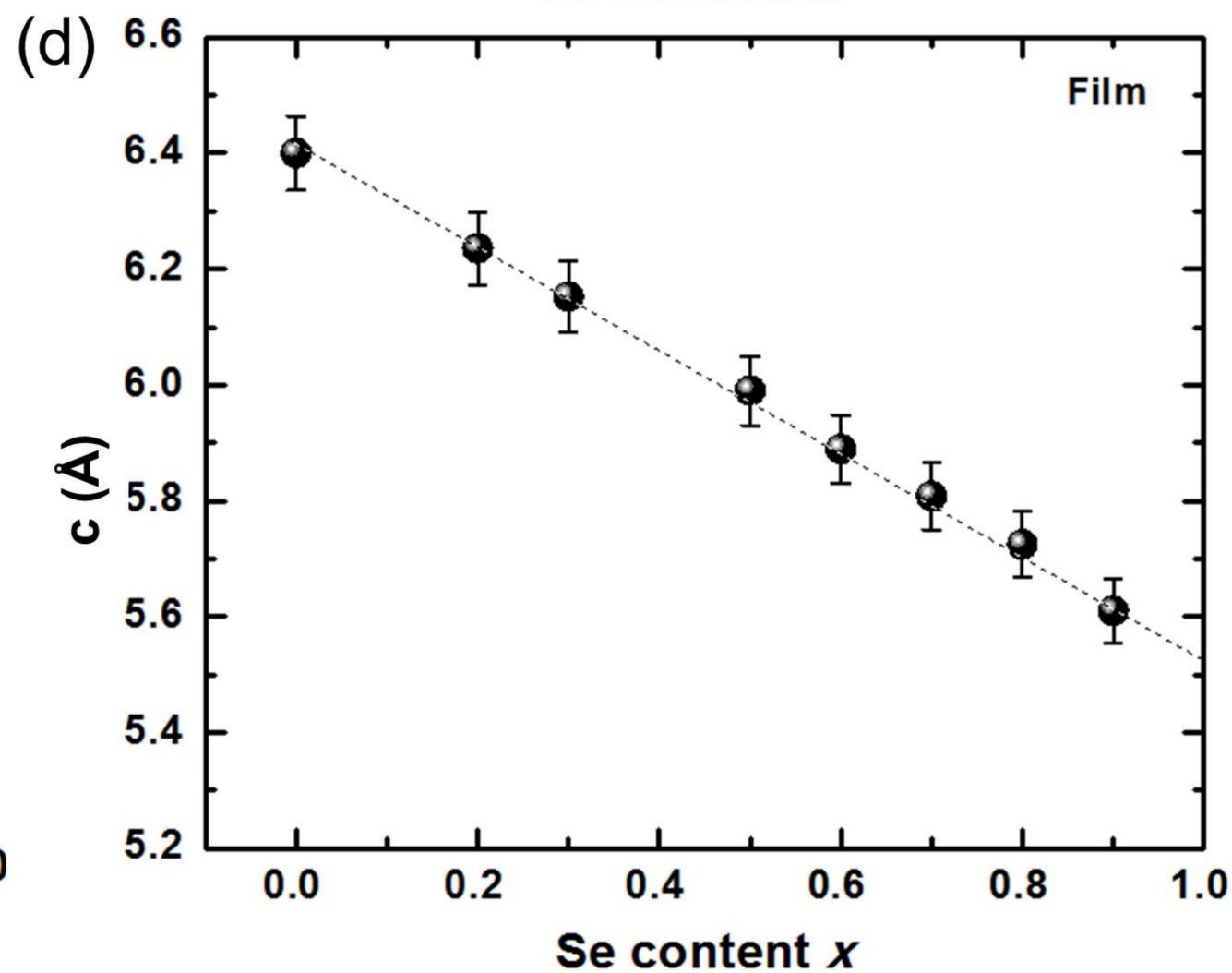

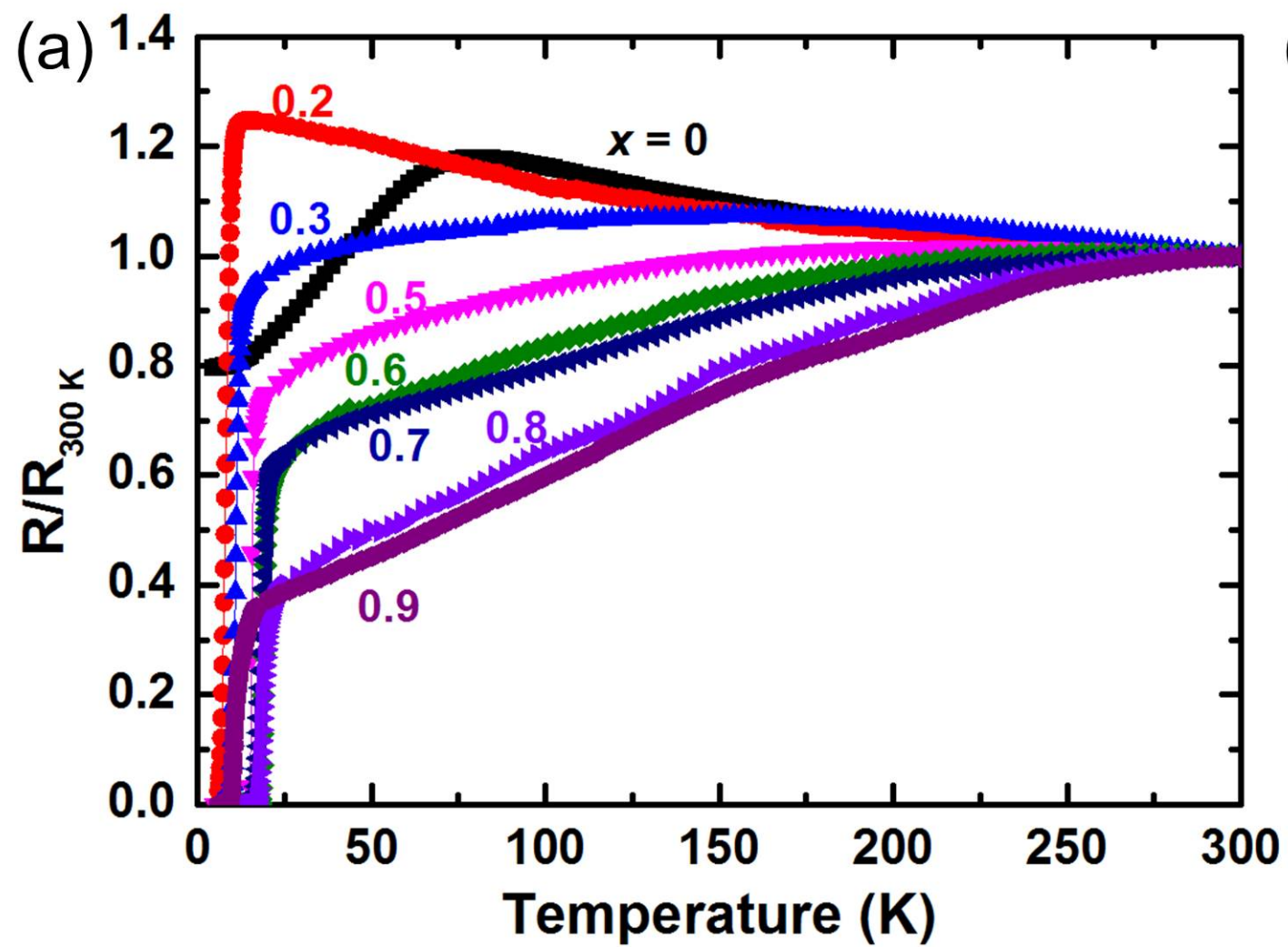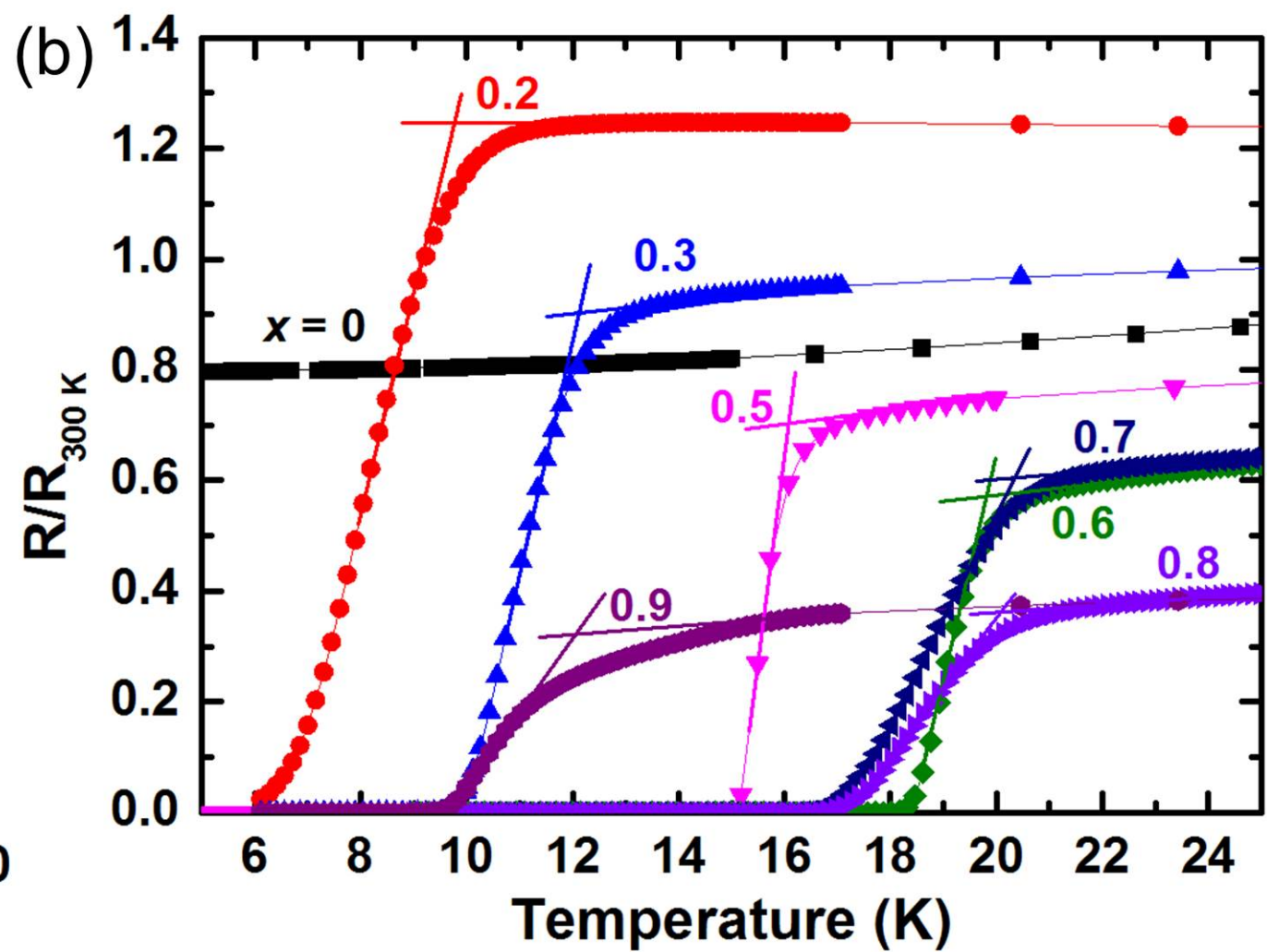

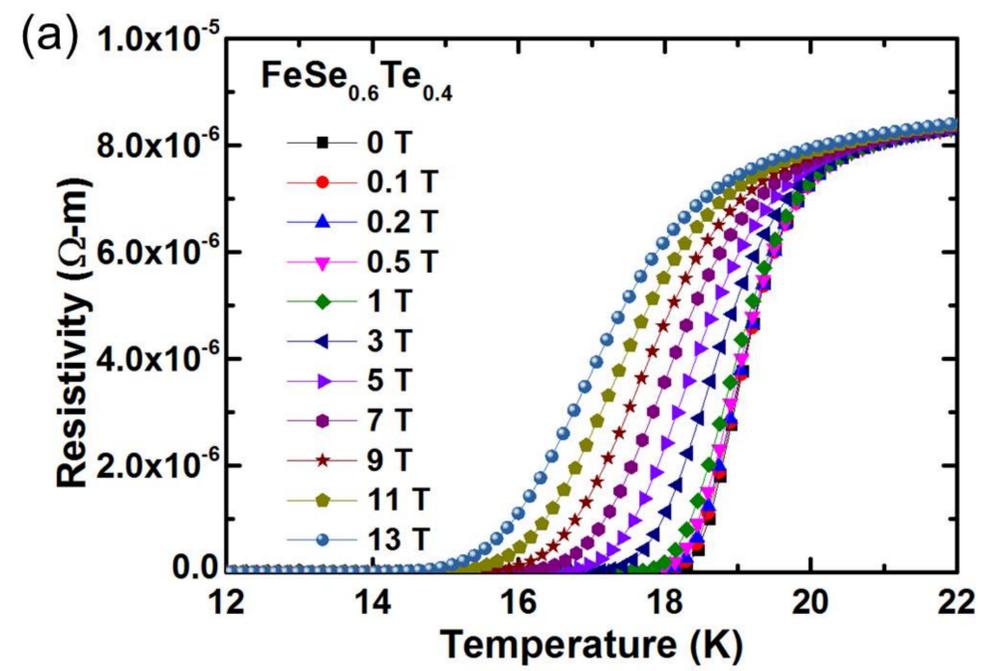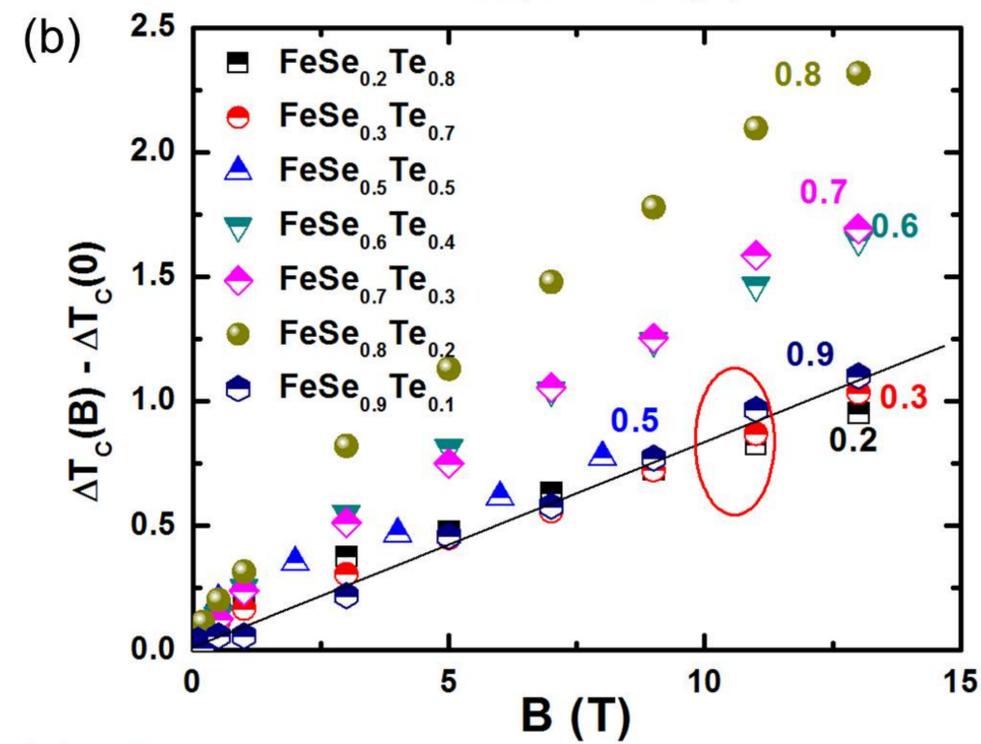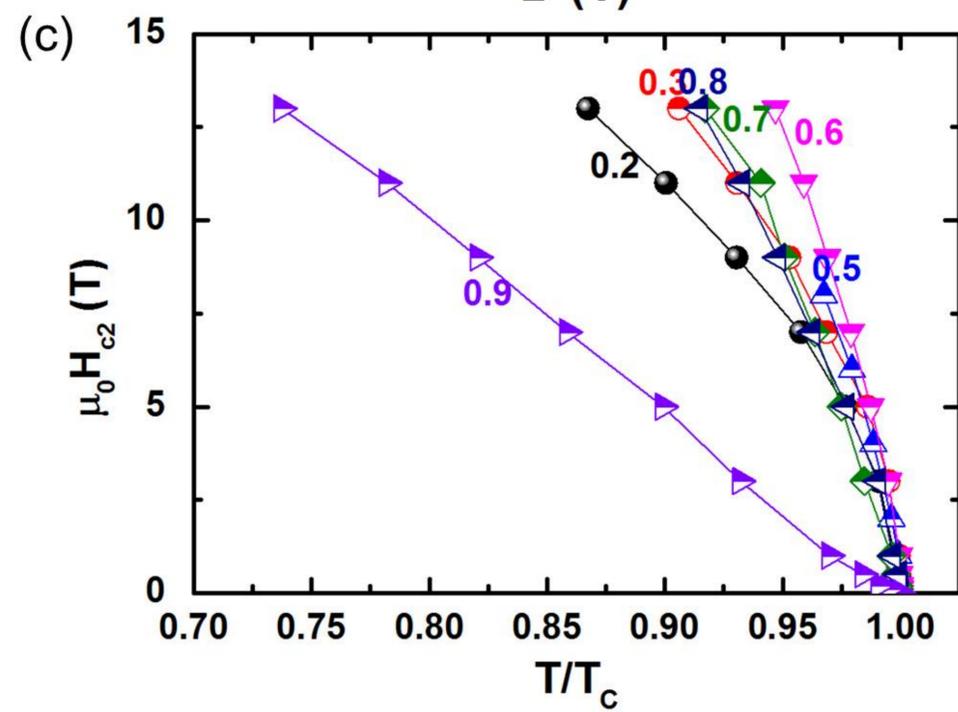

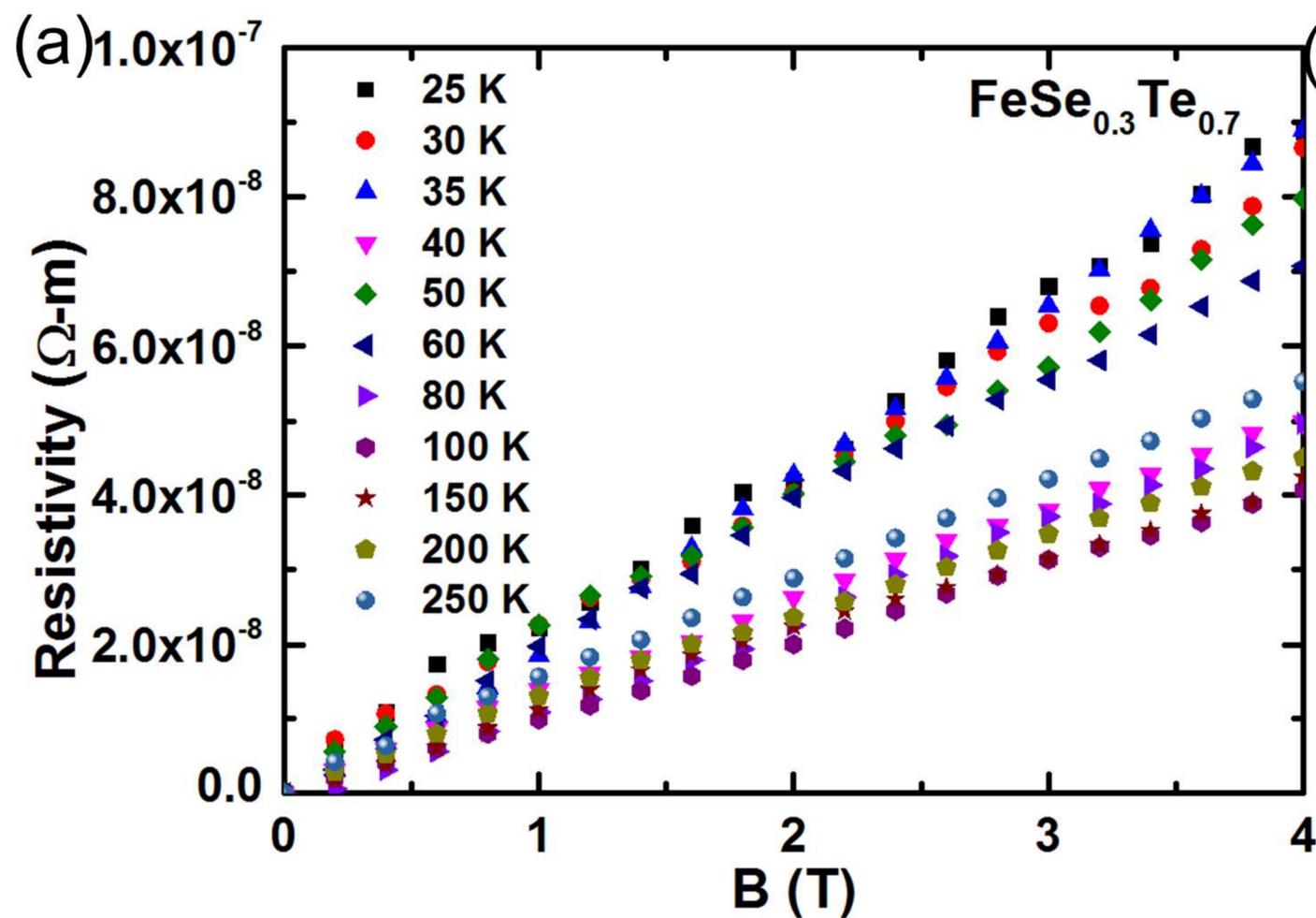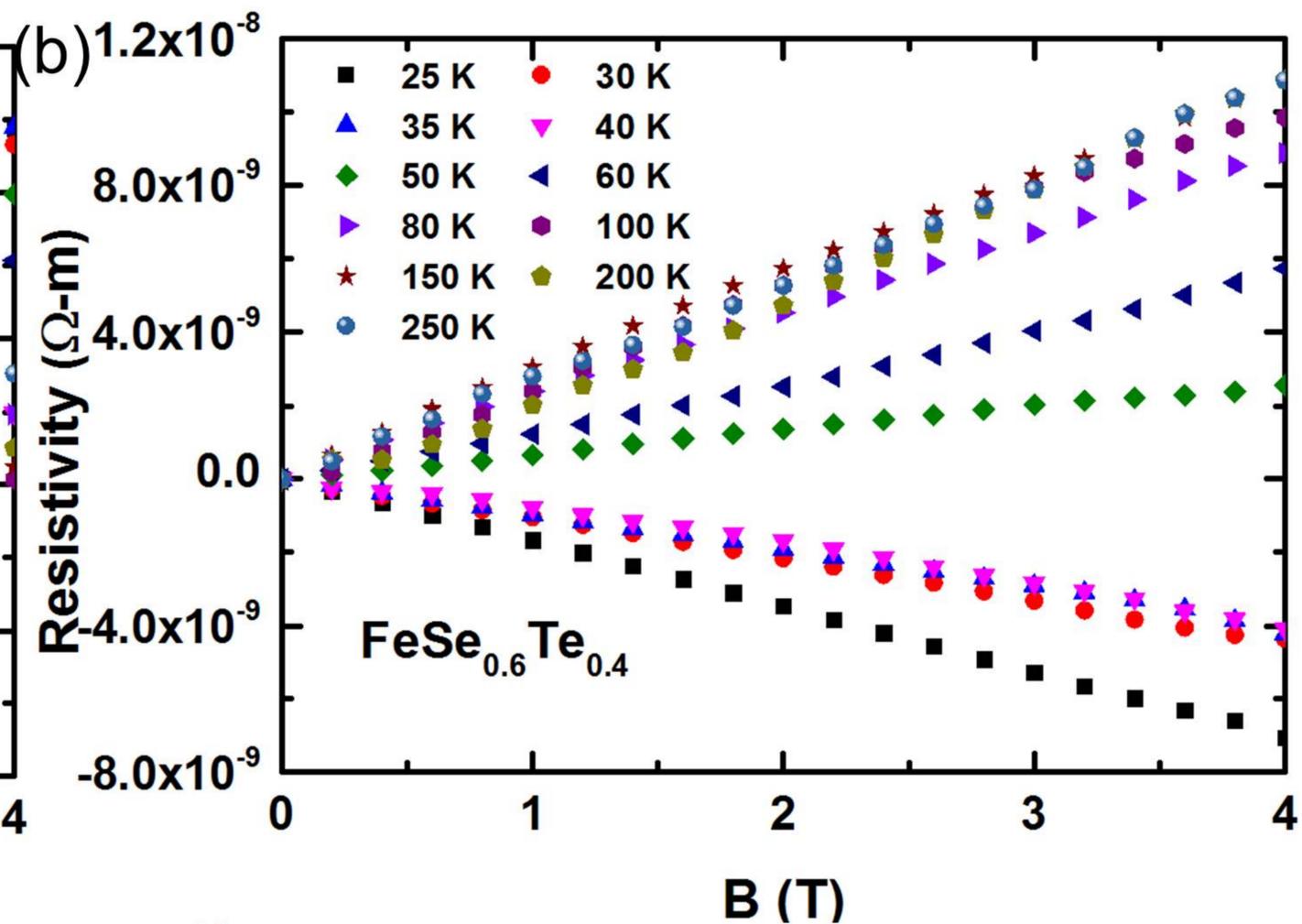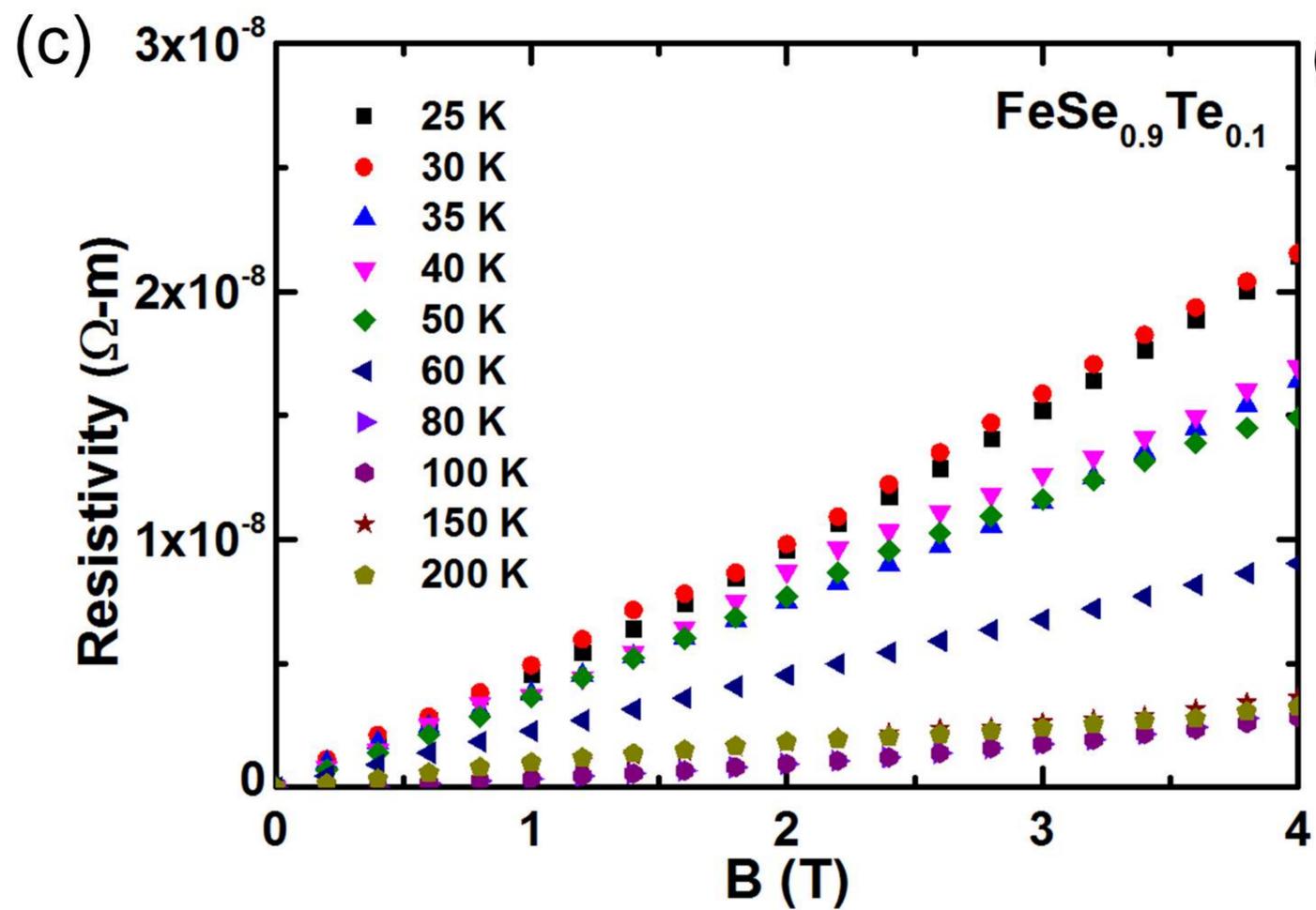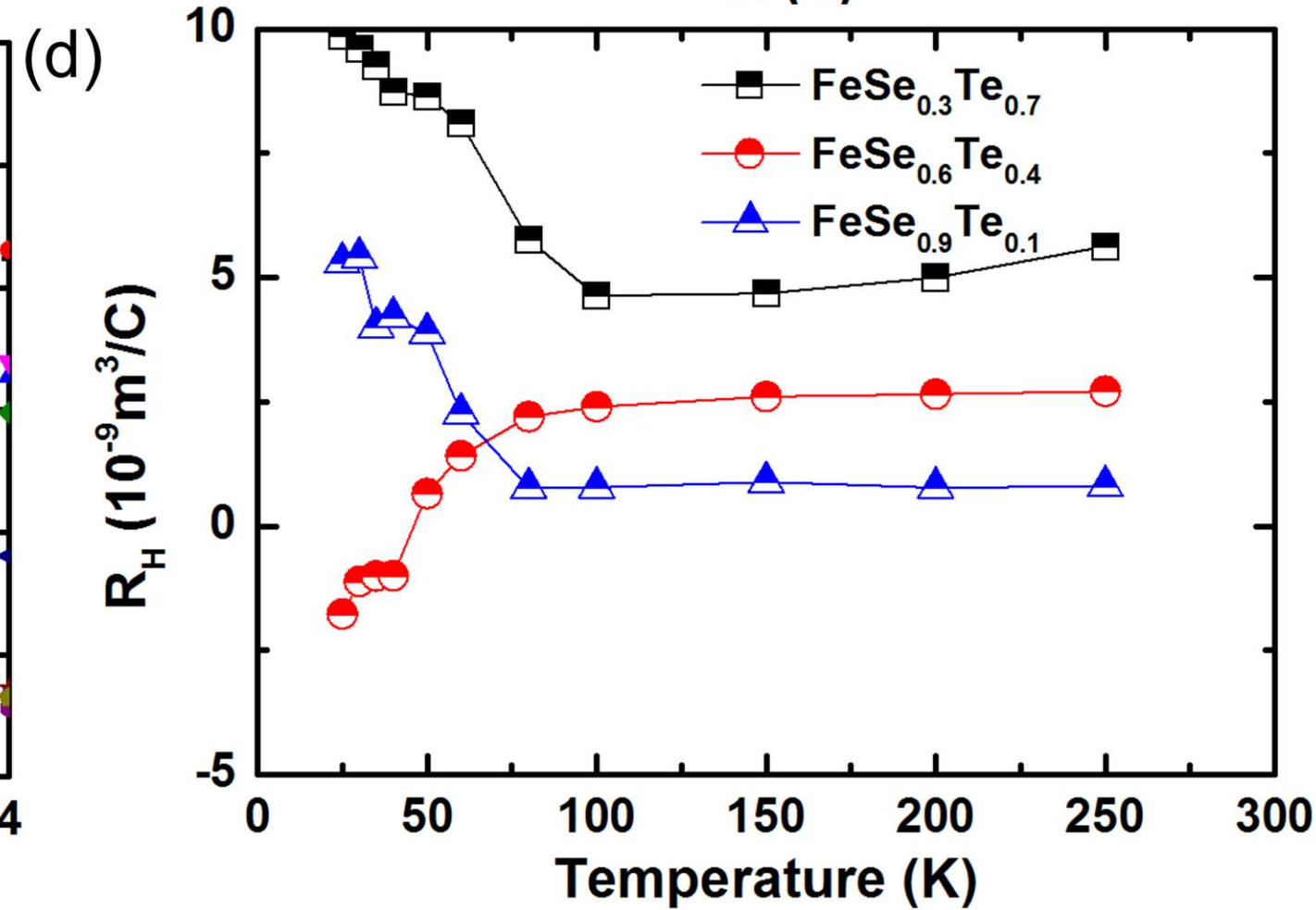

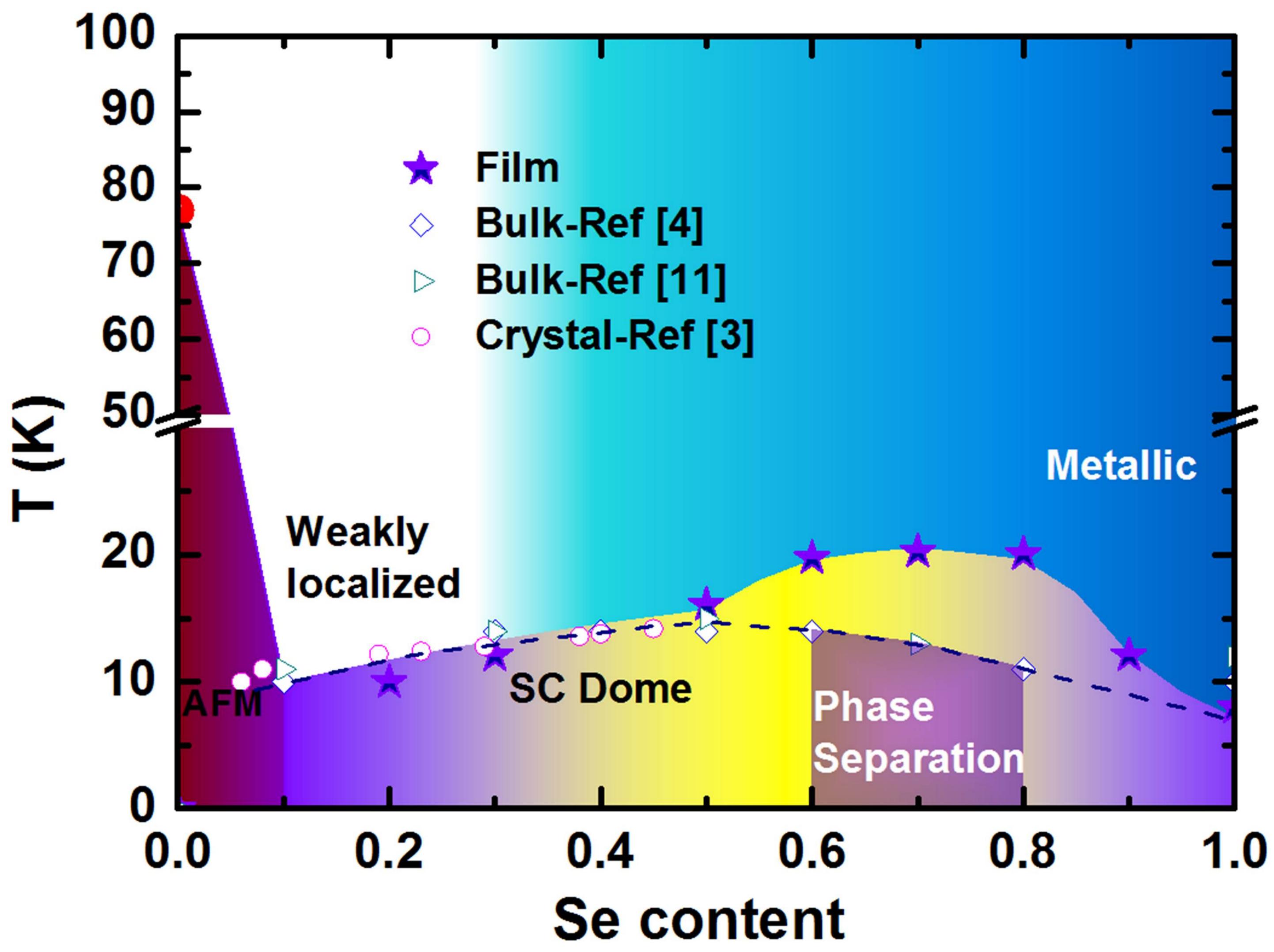